# Government Solvency, Austerity and Fiscal Consolidation in the OECD: A Keynesian Appraisal of Transversality and No Ponzi Game Conditions

Karim AZIZI[1], Nicolas CANRY[2], Jean-Bernard CHATELAIN[3], Bruno TINEL[4]

April 24th, 2013

**Abstract:**

This paper investigates the relevance of the No-Ponzi game condition for public debt (i.e. the public debt growth rate has to be lower than the real interest rate, a necessary assumption for Ricardian equivalence) and of the transversality condition for the GDP growth rate (i.e. the GDP growth rate has to be lower than the real interest rate). First, on the unbalanced panel of 21 countries from 1961 to 2010 available in OECD database, those two conditions were simultaneously validated only for 29% of the cases under examination. Second, those two conditions were more frequent in the 1980s and the 1990s when monetary policies were more restrictive. Third, in tune with the Keynesian view, when the real interest rate is higher than the GDP growth, it corresponds to 75% of the cases of the increases of the debt/GDP ratio but to only 43% of the cases of the decreases of the debt/GDP ratio (fiscal consolidations).

[1] Centre d'économie de la Sorbonne (CES), Université Paris 1 Panthéon Sorbonne,
e-mail Karim.azizi@univ-paris1.fr
[2] Centre d'économie de la Sorbonne (CES), Université Paris 1 Panthéon Sorbonne,
e-mail : Nicolas.canry@univ-paris1.fr
[3] Corresponding author: Centre d'économie de la Sorbonne (CES), Université Paris 1 Panthéon Sorbonne, Paris School of Economics, e-mail Jean-Bernard.chatelain@univ-paris1.fr
[4] Centre d'Economie de la Sorbonne (CES), Université Paris 1 Panthéon Sorbonne,
e-mail : Bruno.tinel@univ-paris1.fr

**Solvabilité des Etats, austérité et consolidation fiscale dans les pays de l'OCDE :**
**Une évaluation Keynésienne des conditions de transversalité et d'absence de jeu à la Ponzi**

**Résumé :**
Le présent article évalue la pertinence de la condition de transversalité sur le taux de croissance du PIB (le taux de croissance du PIB ou du capital doit être inférieur au taux d'intérêt réel) et de la condition d'absence de système de Ponzi pour la dette publique (le taux de croissance de la dette doit être inférieur au taux d'intérêt réel). Cette dernière condition est nécessaire pour prouver l'équivalence Ricardienne. Les deux conditions n'ont été validées simultanément que dans 29% des cas pour l'échantillon non cylindré de 21 pays de 1961 à 2010 disponible auprès de l'OCDE. Elles ont été plus fréquentes lorsque les politiques monétaires sont devenues plus restrictives durant les années 1980 et 1990. En accord avec le point de vue keynésien, lorsque le taux d'intérêt réel est au-dessus du taux de croissance du PIB, il correspond à 75% des cas d'accroissement du ratio de dette publique rapportée au PIB en valeur mais à seulement 43% des cas de diminution du ratio de dette publique rapportée au PIB.

**Mots clés :** Solvabilité des Etats, austérité, consolidation fiscale, condition de jeu à la Ponzi, condition de transversalité, politique budgétaire contra-cyclique, politique monétaire, croissance économique.

*"But this long run is misleading guide to current affairs."* Keynes (1923).

*"The average realized real rate of return on government debt for major OECD countries over the last 30 years has been smaller than the growth rate. Does this imply that governments can play a Ponzi debt game, rolling over their debt without ever increasing taxes?"* Blanchard and Weil (1992).

**1. Introduction**

This paper investigates the relevance of the No-Ponzi game condition for public debt and of the transversality condition for the GDP growth rate, which have been endemic in graduate macroeconomic textbooks for the last twenty years. The No-Ponzi game condition for public debt states that the public debt growth rate has to be lower than the real interest rate. The transversality condition states that the growth rate of GDP (and the growth rate of the capital stock) has to be lower than the real interest rate. Their relevance is assessed first with respect to their ability to describe observed macroeconomic data, and second with respect to a normative macroeconomic policy view which concerns the long term solvency of public debt. Those conditions are necessary hypothesis in order to obtain the theoretical result of Ricardian equivalence. Ricardian equivalence is such that a rise of public expenditures financed by taxation or by public debt does not lead to an inter-temporal change of consumption in the private sector, with zero effect of the Keynesian multiplier of public expenditures on output, because the private sector expects to be taxed later in case of initial public debt. The No Ponzi game conditions are also ruling out bubbles and inefficient capital markets. They are key assumptions at the origin of the current failure of financial macroeconomics (Chatelain and Ralf (2012a)).

Firstly, on the unbalanced panel of 21 countries from 1961 to 2010 available in OECD database, those two conditions were simultaneously validated only for 29% of the cases under examination. Secondly, those two conditions were more frequent in the 1980s and the 1990s when monetary policies were more restrictive. Thirdly, in tune with the Keynesian view, when the real interest rate is higher than the GDP growth, it corresponds to 75% of the cases of the increases of the debt/GDP ratio but to only 43% of the cases of the decreases of the debt/GDP ratio (fiscal consolidations).

As a consequence, those textbooks considered in many economics departments as reference books bias the mind-sets of graduate students (some of them becoming future policy makers or economic advisors) with respect to what really happens in the economy.

The paper is organized as follows. Section 2 discusses several theoretical consequences of the No-Ponzi game and the transversality conditions in current graduate macroeconomic textbooks. The counter-cyclical Keynesian policy point of view with respect to those two conditions is then outlined in section 3. In section 4, a statistical analysis describes the occurrences of the above cases for OECD countries over the last 50 years with a split by decades. A short conclusion follows.

## 2. The No Ponzi Game and the Transversality Conditions, Government Solvency and Ricardian Equivalence

The No-Ponzi Game condition (henceforth the NPG condition) on public and private debt (which also stands for a transversality condition for debt) eliminates the possibility of a Ponzi chain letter by stating that the growth of private debt and of public debt has to be lower than the real interest rate charged on this debt in the infinite horizon. More precisely, let us quote a textbook by

Heijdra and Van Der Ploeg (2002, p.479):

> *"Provided that the agent has **free access to the capital market**, the choice of the problem so far is not meaningful: the agent can simply borrow an infinite amount, service the debt with further borrowings, and live in a state of utmost bliss (presumably that would mean "all fun and no work", with consumption tending to infinity and worked hours to zero). Obviously, something is missing in the story up to now to make for interesting macroeconomics. **The key to the puzzle** is obtained by integrating the dynamic budget equation (the wealth accumulation or the flow of funds equation)."*

$$\frac{\partial a}{\partial t} = (r-n)a_t - (C_t + T_t - W_t) \Rightarrow a_\tau = \int_\tau^{+\infty}(C_t + T_t - W_t)e^{-(r-n)t}dt + \left[\lim_{t\to+\infty} a_t e^{-(r-n)t}\right] \quad (1)$$

where the time index is denoted *t*, *n* is the exogenous growth rate of population, *a* is real financial assets per capita (when negative, it represents the household's debt), *r* is a real interest rate, *W* is the real wage, *T* is lump-sum tax per capita and *C* is consumption per capita of a homogenous good, according to this version of the Ramsey model choice of intertemporal consumption.

Barro and Sala-I-Martin (2004), p.89 and p.92 make a distinction calling the transversality condition on debt (or assets) as the **equality** of the term in square bracket in equation (1) to zero: "*It would be suboptimal for households to accumulate positive assets forever at the rate r or higher, because utility would increase if these assets were instead consumed in finite time*". They define the no Ponzi game as an **inequality** condition such that the term in square brackets is at least positive. With their definition, the transversality condition on debt implies the no Ponzi game condition. Their explanation is as follows: "*In order to borrow on this perpetual basis (schemes in which households' debt grows forever at the rate r or higher), households would have to find willing lenders; that is, other households that were willing to hold positive assets that grew at the rate r or higher. But we already know from the transversality condition that these other households will be unwilling to absorb assets asymptotically at such a high rate*". Besides the assumption of "free access to the capital market on all finite dates", they add the auxiliary No Ponzi game assumption which may be interpreted as "credit rationing in the infinite horizon" as lenders refuse to loan.

When the growth rate of assets is lower than the real interest rate (i.e. when the NPG condition holds), the household intertemporal budget constraint says that the value of financial assets that the agent possesses in a given period must equal the present discounted value of the excess of consumption over after-tax labor income.

The same reasoning applies when adding real government debt (deflated by the GDP deflator) in the Ramsey model (Heijdra and Van Der Ploeg, 2002, p.442). The government identity (in per capita form and omitting seignoriage) is given by a differential equation that could be integrated:

$$\frac{\partial}{\partial t}\left(\frac{D}{p}\right) = (r_t - n)\frac{D_t}{p_t} - \left(\frac{T_t - G_t}{p_t}\right) \Rightarrow \frac{D_\tau}{p_\tau} = \int_\tau^{+\infty}\left(\frac{T_t - G_t}{p_t}\right)e^{-(r_t-n)t}dt + \left[\lim_{t \to +\infty}\frac{D_t}{p_t}e^{-(r_t-n)t}\right] \quad (2)$$

Public debt per capita is denoted $D$, the GDP deflator is denoted $p$, lump-sum taxes per capita are denoted $T$, government consumption per capita is denoted $G$, $r$ is the real interest rate on public debt, t is a time index. The No Ponzi game condition for government is such that:

$$\lim_{t \to +\infty}\frac{D_t}{p_t}e^{-(r_t-n)t} = 0 = \lim_{t \to +\infty}\frac{D_0}{p_0}e^{(g_{D,t}-n)t}e^{-(r_t-n)t} \Rightarrow \lim_{t \to +\infty}g_{D,t} = r_t - \left(\frac{T_t - G_t}{D_t}\right) < \lim_{t \to +\infty}r_t \quad (3)$$

On a given finite date, the growth of real government debt (denoted $g_{D,t}$) is below its real interest rate **only if there is a budget surplus**. The No Ponzi game (NPG) condition assume that this result is valid in the last period of the model (here, the infinite horizon), for limit values of real government debt and its real interest rate. It is then equivalent to state that there is a strictly positive budget surplus in the last period (the infinite horizon). This condition implies in equation (2) that public debt is then exactly equal to the forward looking present value of all the discounted future primary (positive or negative) surpluses. With the NPG condition, it is necessary for the existence of a (positive) public debt that the sum of the cumulated discounted future public surpluses exceeds the sum of the cumulated discounted future public deficits. Under these conditions, there are infinitely many paths for future taxes and futures government expenditures, so that the NPG condition does not exclude future public deficits.

The NPG condition is one of the three necessary assumptions in order to obtain "Ricardian equivalence" in Ramsey models of inter-temporal consumption with discounting. Ricardian equivalence states that budgetary policy has no effect on consumption and the Keynesian fiscal multiplier is equal to zero. For example, a reduction in government saving resulting from a lump sum tax cut is fully offset by higher private savings, so that consumption is unchanged. More precisely, three assumptions are required for Ricardian equivalence (Barro (1974)): (1) Households maximize intertemporally an utility function with aversion with respect to consumption fluctuations without credit constraints, (2) Government finances only government consumption with lump-sum taxes and public debt (no public productive investment, no distortionary taxation, nor money creation), (3) **the assets (or debt) owned by households and by government are exactly zero at the final date** in finite horizon or infinite horizon. The transversality conditions on public and private debt are assumptions (3) for the infinite horizon case. Once (1) and (2) are assumed, there is a mathematical equivalence: Ricardian equivalence holds if and only if the transversality conditions on public and private debt hold.

Let us now turn to the transversality condition on the growth of accumulated capital. It is introduced the neo-classical model of investment, where this time, it is the growth of capital which has to be lower than the real interest rate used as a discount rate in the infinite horizon. It appears also in all the endogenous growth literature with infinitely living agents (for example, in the AK model), where the endogenous growth of output, capital and consumption is limited by a ceiling derived from transversality condition that the balanced growth rate of capital, of output and of consumption has to be lower than the discount rate. The claim appears here that "utility has to be bounded".

Knowing the transversality condition on capital and output, it is now useful to consider its interaction with the dynamics of ratio of government debt to nominal income denoted $pY$ where $Y$ is real gross domestic product (GDP), following Haliassos and Tobin (1990) and still omitting seigniorage:

$$\frac{\partial}{\partial t}\left(\frac{D_t}{p_t Y_t}\right) = (r_t - g_{Y,t})\frac{D_t}{p_t Y_t} - \left(\frac{T_t - G_t}{p_t Y_t}\right) \Rightarrow \frac{D_\tau}{p_\tau Y_\tau} = \int_\tau^{+\infty}\left(\frac{T_t - G_t}{p_t Y_t}\right)e^{-(r-g_Y)t}dt + \left[\lim_{t \to +\infty}\frac{D_t}{p_t Y_t}e^{-(r-g_Y)t}\right]$$

where $g_Y$ is the growth rate of real income. The transversality condition on output implies that government debt to nominal income is positive only when the government will run a cumulated nominal primary surplus on average in the future.

Let us now turn to the issue of **fiscal consolidation** defined as a decrease of the debt over nominal GDP ratio on a given period. This ratio decreases when the growth rate of real debt is lower than the growth rate of real income. The debt dynamics flow of funds equation implies that a **lower bound** for primary deficits or surplus is required for fiscal consolidation:

$$\frac{\partial}{\partial t}\left(\frac{D_t}{p_t Y_t}\right) < 0 \Leftrightarrow g_{D,t} < g_{Y,t} \Leftrightarrow r_t - g_{Y,t} < \frac{T_t - G_t}{D_t} = r_t - g_{D,t}$$

The transversality condition on capital and output $0 < r_t - g_{Y,t}$ implies a destabilizing "snowball effect" which increases the ratio of debt to nominal income. As a consequence, the transversality condition on output and capital requires a **larger** surplus for fiscal consolidation than when this condition does not hold, where relatively small public deficits are compatible with fiscal consolidation because of the reverse stabilizing effect on the ratio of government debt to nominal income when $r_t - g_{Y,t}$ is negative. When the transversality condition on capital and output holds on a given date $0 < r_t - g_{Y,t}$, regimes with deficits ($r_t - g_{Y,t} < 0$: a current period "No Ponzi game condition" does not hold) or regimes with a relatively small surplus lead to an increase of the debt to nominal GDP.

Let us consider a broader model where, for example, the growth rate of output depends negatively on the real interest rate, and where the budget surplus (in particular tax income) is an increasing function of the growth rate of output and then a negative function of real interest rates. In this case, when the transversality condition on capital holds on a given period, one may observe frequently public deficits, which means a rejection of the current period equivalent of the No Ponzi Game condition. Hence, it may not be so frequent that both current period equivalents of the transversality conditions on capital and on public debt hold simultaneously, as seen in section 4.

What is more, the transversality conditions are built on shaky mathematical and economic grounds for the following reasons.

1. It is the choice of a terminal condition for infinite horizon problems which is necessary only when lifetime utility is finite at the optimum (Kamihigashi, 2005). By analogy to a finite horizon terminal condition, it states that the discounted value of capital in the infinite horizon is zero. The Halkin (1974) counter-example demonstrates that *in general*, there are no necessary transversality conditions for infinite horizon optimal control problems when one does not assume that the objective function converges. Moreover, even when the objective function does converge in Halkin's (1974) counterexample, Caputo (2005, chapter 14) still concludes it is a valid counterexample for demonstrating that the usual textbook transversality condition is not necessary, contrary to the claim of Chiang (1992, Chapter 9). The first model of this type was proposed by Ramsey (1928) with an objective function without discounting. Ramsey did not assume those transversality conditions, and his model is still not considered as flawed.

2. The households maximising utility prefers infinite utility with respect to bounded utility. For this reason, it does not matter to keep non-zero positive wealth in the infinity limit (see an example in Amable, Chatelain, Ralf (2010)).

3. The agents (households, firms, government) may **not** have free unlimited access to the capital market at all finite future dates, which is another key to the puzzle put forward in Heijdra and Van der Ploeg quote. Amable, Chatelain and Ralf (2010) for private debt and Chatelain and Ralf (2012a) for public debt propose *alternative solvency "collateral" constraints* on all future finite dates set by imperfect capital market in Ramsey models of infinitely living agents with discounting. Setting credit constraints and covenants at each date for the next finite period is a more natural way to fight against Ponzi behaviour than setting an infinite horizon zero credit constraint such as the No Ponzi Game condition. Imagine that we apply a similar reasoning as in the financial accelerator: public debt is solvent based on the capacity to repay bonds based on next period expected distortionary taxes on output net of public expenditures. And let us assume the lenders expect it to hold for all future finite dates $t$:

$$(1+r_t) \cdot B_t < \tau_{t+1} \cdot Y_t(1+g_{Y,t+1}) - G_t(1+g_{G,t+1}) \Rightarrow B_t < [\tau_{t+1} \cdot Y_t - G_t(1+g_{G,t+1} - g_{Y,t+1})](1+g_{Y,t+1} - r_t)$$

The second inequality is a first order Taylor development of the first inequality. Government is solvent because debt is backed by the discounted next period surplus in a similar fashion than equation (2), but, now, the transversality conditions on capital and on debt **related to a relatively large real interest rate** are **no longer required** in order to find this statement: a solvent government debt needs to be backed by discounted future surplus. In particular, if the interest rate on public bonds is low and if the expected growth in output is large (which is exactly the **opposite** of the infinite horizon transversality condition on output or capital), this solvency constraint is more likely to be respected. This is what bond holders may think about solvency in the short run, and they may not care that government should hold zero debt at the end of times. They would like taxes to increase and public expenditures to fall, but would enjoy growth in output even more, as it increases the taxable base.

Imagine that this collateral constraint holds **all** future periods. Then government debt is always solvent. Imagine that at the same time, the growth rate of output is equal to the growth rate of public debt, but is larger than the real interest rate on public debt. Then, the infinite horizon (i.e. The NPG) solvency constraint is not fulfilled, whereas the short run solvency constraint is always fulfilled. In this case, the infinite horizon solvency constraint is meaningless. In this context, government and the private sector may hold strictly positive or negative assets in the last period of the model (the infinite horizon), so that **Ricardian equivalence does not hold, the Keynesian multiplier has no zero effect and government is solvent.**

If one introduces uncertainty in the above setting, then a key issue for solvency may be related to the investors' time horizon for the expected growth rate of the country. If they take into account the expected growth rate for the next ten years, solvency problems are very likely to be minimal, even when adding uncertainty related to the growth of output. But, if they take into account only the short run (next year's growth rate), they may over-lend and suddenly stop a few years later, quickly leaving this country's sovereign bonds market. In this case, an additional simultaneous equation is required, where the risk premium determining the interest rate of public debt depends on the probability of default which is related to the above equation.

Finally, these solvency and collateral capital constraints make for much more interesting macroeconomics than the infinite horizon solvency of the No-Ponzi game condition in macroeconomics for at least three reasons:

1. The Keynesian multiplier may be effective, which is not the case with Ricardian equivalence.

2. The No-Ponzi Game condition related to private agents rules bubbles of private assets out of the model. This assumption is used to rule out *the existence* of bubbles *prior to the infinite horizon* in macroeconomic models. This is consistent with the efficient financial market hypothesis. But this is an issue in dealing with financial crises and the link between monetary and macro-prudential policy (Chatelain and Ralf, 2012a).
3. In the endogenous growth literature, the transversality conditions are inconsistent with growth miracles. For growth miracles, the growth of output consistently exceeds the real rate of interest for several decades as was the case for Japan between 1960 and 1990 or China between 1990 and 2010 (Amable, Chatelain, Ralf, 2010).

Although the NPG and transversality conditions are stated for the "long run" infinite horizon, several macroeconomic models consider a fixed value of the interest rate, so that this NPG condition holds for all periods, and it is not only a limit condition. In real data, any time can be the short run "now" and the long run of many years ago. Hence, the next section investigates short run properties of the inequalities related to the NPG condition and the transversality condition on output in pre-NPG conditions, namely, in non-modern undergraduate macroeconomics such as IS-LM type Keynesian macroeconomics.

## 3. Countercyclical Keynesian Policies and the No Ponzi Game and Transversality Conditions World

In contrast to the infinite horizon No Ponzi Game and Transversality conditions approach, the Keynesian view takes into account the short run position of the economy in the business cycle. As the instability of the economic system is acknowledged, Keynesian monetary and fiscal policy prescriptions are liable to be counter-cyclical. In line with early Keynesian authors such as Hansen and Greer (1942), Lerner (1943) and Domar (1944), public debt (defined as the sum of accumulated deficits) is first understood on the ground of the fiscal multiplier. Ricardian equivalence is not considered relevant so that the fiscal multiplier is not equal to zero. If the economy does not use its full production capacity, then an increase in public spending will induce faster growth, since production is supposed to be demand-led. Tax cuts are said to induce the same type of adjustments but with less intensity and their impact on public debt dynamics can be somewhat different, but we do not examine this issue further in the present paper (Pucci and Tinel, 2011). Keynesian economists generally acknowledge that such demand increases which occur through public outlays should give rise to the highest possible multiplier effect if financed by debt rather than taxes.

For several Keynesian economists, as long as the system is not at full employment, crowding out effects should be negligible (Arestis and Sawyer, 2004a, 2004b). From this point of view, public debt does not really compete with the supply of private assets because savings are endogenous with respect to public spending. A saving level is induced by public spending through national income adjustments. For this reason, long term interest rates are not supposed to rise mechanically with public debt. Besides, in the case of underemployment, public bonds and private assets are not competing against each other for funds because the requirements of portfolio diversification make them much more complementary than substitutable, as they bear different yields and risks.

This presentation of discretionary fiscal policies needs a few additional comments relating to automatic stabilisers. When growth accelerates, public spending automatically slows down because less urgent public spending is required to aid people in facing unemployment and poverty, meanwhile more taxes are levied on revenues and transactions simply because these are increasing. As a result, public deficit and public debt to GDP ratio are automatically reduced with more growth. When growth is slowing down, the opposite result is obtained: more public deficit and higher public debt to GDP ratio. A typical Keynesian idea is that even so called automatic stabilisers are not sufficient to improve economic activity suitably (i.e. to reach the level at which is starts to create jobs again); if no discretionary expansionist policy is undertaken, then the economy is likely to remain locked much longer in a situation where the debt/GDP ratio increases ($g_D > g_Y$). Though such a situation seems to have persisted over time in Europe, during the last 30 years, it appears that governments in fact resorted to countercyclical (Keynesian) discretionary fiscal policies (Amable and Azizi, 2011).

Once growth has been stimulated through public investment and/or final consumption, the resulting increase in national income leads in turn to an increase in tax receipts. At the end of the process (in the long run), if the size of the multiplier is greater than 1 during recessions (which is often assumed under reasonable hypotheses but not always empirically verified, see the recent new evaluations by Blanchard and Leigh (2013) for the current crisis), the rise in output is expected to be more important than the rise in public outlays ($g_Y > g_G$) and the rise in tax receipts $g_T > 0$ is supposed to compensate at least partially for the initial additional public spending which reduces both public deficit and debt. Note that this result is likely not to be observable instantaneously or on a very short period of time because of time lag and multiplier time processes.

Then, if the real interest rate is "not too high" (lower than the GDP growth rate), the ratio of public debt to GDP is supposed to be smaller at the end of the process than at its beginning. Though the level of public debt is higher, it is compensated for by an even higher level of GDP. In other words, the growth rate of the nominal public debt measured on the whole process is expected to be smaller than the growth rate of the domestic revenue during the same period of time: $g_D < g_Y$. Of course, this result depends heavily on the elasticity of tax receipts to growth and has to be amended if the real interest rate $r$ at which the government is able to issue bonds is greater than $g_Y$. In this situation, the "snowball" effect implies that the government has to run a primary surplus just to stabilise its debt to GDP ratio. In a macroeconomic context where the condition $r > g_Y$ holds, any deficit spending leads to $g_D > g_Y$.

During booms, the fiscal multiplier mechanism can be used the other way round: a government can run public surpluses in order to reduce demand and hence limit the GDP growth rate if the economy is already at full-employment. Such a policy reduces the debt to GDP ratio. The government has to behave in a countercyclical way: if the growth of output $g_Y$ is low (when private demand is not sufficient to improve the level of employment) then deficit spending, which increases the debt over GDP ratio ($g_D > g_Y$) in the short run; if the growth of output $g_Y$ is high (when private demand is sufficient to induce a reduction in unemployment) then running surpluses which reduces the debt over GDP ratio ($g_D < g_Y$) in the short run.

Besides counter-cyclical budgetary policy, a Keynesian monetary policy consists in keeping interest rates low enough as economic activity slows down to prevent the cost of private investment from being too high and too much of a deterrent to investment and also to keep the cost of public debt as low as possible. Raising real interest rates above the GDP growth rate for a prolonged period when the economy is not at full-employment is clearly not a good monetary policy prescription from a Keynesian point of view. In this framework, many Keynesian economists would consider it preferable to give priority to employment even when $r$ is "high", i.e. $r > g_Y$. In other words, as long as full-employment is not realised, $g_D > g_Y$ is expected even when $r > g_Y$. The "snowball" effect cannot be considered as a deterrent factor to deficit spending for a Keynesian government as long as full-employment is not attained.

Moreover, if ever the NPG and the transversality conditions were followed by a government then the Keynesian view contends that it would be likely not to lead to the result claimed by its proponents. As explained in section 2, public debt will increase when there is a nominal primary deficit because of a "snowball" effect, that is, when the growth rate of output is below the average interest rate on government bonds. The current period equivalent of the NPG and transversality conditions may not hold simultaneously. This doesn't mean that it cannot happen sometimes (probably most of the time just before large recessions).

To some extent, it is possible to specify the behaviour of government according to the macroeconomic situation which is simply characterised by the level of growth and the order of $r$, $g_D$ and $g_Y$. In table 1, we map into several regimes the concordance or the discordance of counter-cyclical policies with the No Ponzi Game and Transversality conditions.

*Table 1: Specification of the policy mix behaviour according to booms or recessions*

| Regime | Conditions<br>561 observations | $g_Y$ is low (recession)<br>$g_Y$ <2.64% (280 observations; 50%) | $g_Y$ is high (boom)<br>$g_Y$ >2.64% (281 observations, 50%) |
|---|---|---|---|
| 1 | Low real interest rate<br>$g_D > g_Y > r$<br>70/561=12% | Deficit, Keynesian Budgetary Policy<br>Expansionary Monetary Policy<br>28/280=10% | Deficit, Pro-cyclical budgetary policy<br>Expansionary Monetary Policy<br>42/281=15% |
| 2 | Medium interest rate<br>$g_D > r > g_Y$<br>175/561=31% | Deficit, Keynesian Budgetary Policy<br>Mild restrictive Monetary Policy<br>134/280=48% | Deficit, Pro-cyclical budgetary policy<br>Mild restrictive Monetary Policy<br>41/281=15% |
| 3 | Transversality, NPG<br>$r > g_D > g_Y$<br>46/561=9% | Surplus, Keynesian Budgetary Policy<br>Restrictive Monetary Policy<br>35/280=13% | Surplus, Pro-cyclical budgetary policy<br>Restrictive Monetary Policy<br>11/281=4% |
| 4 | Low real interest rate<br>$g_Y > g_D > r$<br>41/561=7% | Deficit, Pro-cyclical budgetary policy<br>Expansionary Monetary Policy<br>6/280=2% | Deficit, Keynesian Budgetary Policy<br>Expansionary Monetary Policy<br>35/281=12% |
| 5 | Medium interest rate<br>$g_Y > r > g_D$<br>115/561=20% | Surplus, Pro-cyclical budgetary policy<br>Mild expansionary Monetary Policy<br>15/280=5% | Deficit, Keynesian Budgetary Policy<br>Mild expansionary Monetary Policy<br>100/281=36% |
| 6 | Transversality, NPG<br>$r > g_Y > g_D$<br>114/561=21% | Surplus, <u>Austerity</u>: pro-cyclical budgetary, restrictive monetary policies<br>62/280=22% | Surplus, Keynesian Budgetary Policy<br>Restrictive Monetary Policy<br>52/281=19% |
| 2+3+6 | $r > g_Y$<br>335/561=60% | Snowball effect, During recessions<br>231/280=82% | Snowball effect, During booms<br>104/281=37% |
| 5+3+6 | $r > g_D$<br>275/561=49% | Surplus, During recessions<br>112/280=40% | Surplus, During booms<br>163/281=58% |
| 3+6 | Transversality, NPG<br>160/561=29% | During recessions<br>97/280=35% | During booms<br>63/281=22% |
| 1+2+3 | Debt/GDP increases<br>292/561=52% | During recessions<br>198/280=71% | During booms<br>94/281=33% |

In columns 3 and 4, "surplus" corresponds to $g_D > r$. Fiscal Consolidation or a decrease of the debt/GDP ratio ($g_Y > g_D$) is related to "counter-cyclical budgetary policy" during booms and "pro-cyclical policy" or austerity during recessions. An increase of the debt/GDP ratio ($g_D > g_Y$) is related to counter-cyclical Keynesian budgetary policy" during recessions or "pro-cyclical budgetary policy" during booms. The case where $g_Y > r$ is labeled as "snowball effect" or "expansionary monetary policy" (and the opposite case to "restrictive monetary policy"). We are well aware that this "label" may lead to confusion, firstly, because there are usually inversions of the yield curve before recessions, where the long term government bonds rates (denoted $r$) are lower than the daily monetary policy rates, and, secondly, because expansionary or restrictive monetary are usually defined with respect to output gaps and/or to the wedges between expected inflation or the expected growth of monetary aggregates with respect to their targets. Observed frequencies are discussed in the next section.

**4. Confronting the transversality and No-Ponzi Game conditions with OECD data**

**4.1. The OECD Gross Public Debt Data Set**

In this section, we assess the frequencies of transversality conditions using the currently available OECD database for gross public debt, which corresponds to an unbalanced panel of 21 countries since 1961 to 2010 included (561 observations). The data for the 21 countries enter gradually into the sample: 1961: Netherlands, USA, 1962: Canada, 1967, UK, 1970: Belgium, France, 1981: Spain, 1985: Norway, 1987: Danmark, Sweden, 1989: Japan, 1990: Australia, Austria, 1991: Switzerland, 1992: Germany, Italy, 1996: Portugal, 1998: Greece, 1999: Iceland, Ireland, 2003: Korea. Hence, there are 50 observations for USA and the Netherlands and only 8 observations for Korea. The 10 years government bond yield and the OECD gross public debt are net of the GDP deflator (hence, we may use in what follows the wording: the growth rate of "real public debt"). In the second column of table 1 are presented breakdowns for the overall OECD data set in the six regimes. Breakdowns during booms and recessions are also presented in columns 3 and 4 of table 1. Recessions are roughly defined for country and years where the GDP growth rate is below the overall median of growth rates, equal to 2.64%. It is a brute force classification of booms and recessions knowing for example that structural breaks on growth trends occurred for the five countries for which we had around 46 observations before 1974. This suggests that readers may reasonably consider a margin of error of 20% for proportions found in each cell. However, as we use only medians and order statistics, our modest approach is at least robust to outliers. This may not necessarily be the case of papers publishing bold economic policy statements driven by a few outliers (Chatelain (2010), Chatelain and Ralf (2012b), Herndon, Ash and Pollin (2013)).

**4.2. The No Ponzi Game and Transversality Conditions in Booms versus Recessions**

Over the year 1961 to 2010 years, the two conditions (i.e. the NPG and the transversality conditions) were validated simultaneously on OECD data in only 29% of the cases (table 1, regimes 3 and 6). As seen in the next section, this proportion of cases is biased upwards because of a selection bias with a smaller number of observations for the 1960s and the 1970s with respect to the following decades. In fact, taking into account separately each of the conditions leads to high frequencies (60% for the condition $r > g_Y$ and 49% for the condition $r > g_D$) but it turns out that the high frequencies of the discordant regimes 2 (31% of the cases) and 5 (20% of the cases) confirms that it is difficult to hold together these two conditions because of the "snowball effect".

The transversality condition on output ($r > g_Y$ roughly related to "restrictive monetary policy") is highly frequent during recessions (82% of the recessions) and much less frequent during booms (37% of the cases). By contrast, the No Ponzi game condition on public debt ($r > g_D$) is less frequent during recessions (40% of the recessions) than during booms (58% of the cases). When both condition holds (which removes highly frequent regime 2 in recessions (48%) and highly frequent regime 5 in booms (36%)), it depicts a world which is more frequently in recessions (61% (=97/(97+63))) than in the overall data set (50%). As a consequence, those "reference" textbooks based on both conditions distorts the judgement of graduate students presenting as a "normal case" for the economy two joined conditions **which are more frequently related to recessions**.

### 4.3. Government Solvency and Fiscal Consolidation in Booms versus Recessions

Keynesian counter-cyclical budgetary policy related to an increase of debt/GDP ratio occurs in 71% of recession cases and related to a fiscal consolidation (a decrease of debt/GDP ratio) in 67% of boom cases (table 1, columns 3 and 4). Conversely, fiscal consolidation during recessions ("austerity") occurs in 29% of recession cases and pro-cyclical expansionary budgetary policy during booms occurs in 33% of boom cases.

During recessions, regime 2 ($g_D > r > g_Y$) is the most frequent regime (48% of the cases). Monetary policy may be mildly restrictive, and budgetary policy is Keynesian. Then the transversality condition/No Ponzi game regime 6 ($r > g_Y > g_D$) corresponds to 22% of the cases, running surplus during recessions along with a restrictive monetary policy (the "austerity" regime).

During booms, regime 5 ($g_Y > r > g_D$) is the most frequent regime (36% of the cases). Monetary policy may be mildly expansionary and there is a counter-cyclical fiscal consolidation with a relatively small deficit. Then the transversality condition/No Ponzi game regime 6 ($r > g_Y > g_D$) corresponds to 19% of the cases. Monetary policy may be mildly expansionary and there is a counter-cyclical fiscal consolidation with a budget surplus.

In both booms and recessions, the destabilizing snowball effect (related to the transversality condition on output) implies that the increase in the debt to GDP ratio (table 1, regimes 1, 2 and 3) is strongly tied to cases where the real interest rate is larger than the GDP growth (table 1, regimes 2 and 3) which represent 75% (222/292) of the occurrences of debt to GDP ratio increases (regimes 1, 2 and 3).

By contrast, debt consolidation, which is a signal of an increased solvency of government, is not strongly tied to the NPG and transversality conditions: regime 6 characterizes only 43% of the set of consolidation occurrences (regimes 3, 4 and 6). The remaining 57% of the consolidation situations corresponds to cases where the real interest rate is lower than the GDP growth (regime 4 and 5).

In tune with the Keynesian view, when the real interest rate is higher than the GDP growth, it corresponds to 75% of the cases of the increases of the debt/GDP ratio but to only 43% of the cases of the decreases of the debt/GDP ratio (fiscal consolidations). As a consequence, those "reference" textbooks based on both conditions also distort the judgement of graduate students **with respect to their alleged normative property for the long term solvency of government**.

### 4.4. The No Ponzi Game and Transversality Conditions per Decades

Table 2 presents the median values of the three key variables of interest over the last five decades, which signals marked contrasts (median values are in parenthesis in what follows). In the 1960s, there was high growth (5%), low real interest rate (2.3%) and consolidation of real public debt (2%). In the 1970s, oil shocks and high inflation led to very low real interest rate (0.9%), low growth of real public debt (2.6%) and still higher growth (3.7%), hence a median consolidation of public debt/nominal output ratio. In the 1980s, monetary policy targeted disinflation, with high real rate (median value 5.9%), high growth of real public debt (5.1%) and low growth of output (2.6%), with a large rise of public debt/nominal output ratio. In the 1990s, this went on again with monetary policy still targeting disinflation (from 6% to 2% in the USA and Euro-area countries), with lower real interest rate (4.5%) and higher growth for some countries with the internet bubble (2.9%) with a median value of the growth of debt of 3.6% knowing that the skewness of the growth of debt increased during this decade (the mean is 4.8%). In the 2000s, the great moderation with low real interest rate which followed the burst of the internet bubble, and then the monetary policy close to the zero lower bound for nominal interest rate along with quantitative easing during the second great depression led to a balanced case for median value, with the real interest rate 2% is below the median growth rate 2.1% with is below the growth rate of real public debt (still with a highly skewed distribution, including the outlier Iceland in 2008, with a growth rate higher than 100%).

With respect to Piketty (2011) who investigated the gap ($r > g_Y$) over the whole period 1979-2009 for its effect on bequests in France, our breakdown per decade highlights the downward shift of the rate on returns on bonds which occurred in the 2000s. For the next decade (2011-2020), OECD countries are very likely to remain in a Japanese style liquidity trap regime with low rate of return on bonds close to the level of GDP growth.

*Table 2: Median values of GDP growth rate, of gross public debt growth rate and of average real 10 years government bonds yields (561 OECD annual observations)*

| 1961-1970 | $g_D$ = 2.0 % < **_r = 2.3 %_** < $g_Y$ = 5.0 % |
|---|---|
| 1971-1980 | **_r = 0.9 %_** < $g_D$ = 1.6 % < $g_Y$ = 3.7 % |
| 1981-1990 | $g_Y$ = 2.6 % < $g_D$ = 5.1 % < **_r = 5.9 %_** |
| 1991-2000 | $g_Y$ = 2.9 % < $g_D$ = 3.6 % (mean 4.8%) < **_r = 4.5 %_** |
| 2001-2010 | **_r = 2.0 %_** < $g_Y$ = 2.1 % < $g_D$ = 2.5 % (mean 4.8%) |

*Table 3: Breakdowns of regimes per decade (frequency and column percentage)*

| Regime | Decade | | | | | |
|---|---|---|---|---|---|---|
| | 61-70 | 71-80 | 81-90 | 91-00 | 01-10 | Total |
| 1. Low real interest rate $g_D > g_Y > r$ | **4** **11%** | **15** **25%** | 1 1% | 7 4% | **43** **21%** | 70 12% |
| 2. Medium interest rate $g_D > r > g_Y$ | 2 6% | **7** **12%** | **36** **41%** | **69** **41%** | **61** **29%** | 175 31% |
| 3. Transversality, NPG $r > g_D > g_Y$ | 2 6% | 1 2% | **16** **18%** | **19** **11%** | 8 4% | 46 9% |
| 4. Low real interest rate $g_Y > g_D > r$ | **11** **31%** | **17** **28%** | 0 0% | 3 2% | 10 5% | 41 7% |
| 5. Medium interest rate $g_Y > r > g_D$ | **15** **43%** | **18** **30%** | 4 5% | **25** **15%** | **53** **25%** | 115 20% |
| 6. Transversality, NPG $r > g_Y > g_D$ | 1 3% | 2 3% | **31** **35%** | **47** **28%** | **33** **16%** | 114 20% |
| Total | 35 6% | 60 11% | 88 16% | 170 30% | 208 37% | 561 100% |
| $r > g_Y$ (regimes 2+3+6), snowball effect | 15% | 17% | 90% | 80% | 49% | 60% |
| $r > g_D$ (regimes 3+5+6), budget surplus | 52% | 35% | 40% | 54% | 45% | 49% |
| $r > \max(g_D, g_Y)$ (regimes 3+6) | 9% | 5% | 53% | 39% | 20% | 29% |
| $g_D > g_Y$ (regimes 1+2+3) | 23% | 39% | 60% | 56% | 54% | 52% |

According to table 3, the empirical relevance of the NPG and the transversality conditions varied over time with respect to monetary and budgetary policy changes. The transversality condition $r > g_Y$ had a low frequency in the 1960s and the 1970s (at most 17%), whereas it rose to 90% and 80% in the 1980s and 1990s respectively. Finally, it corresponds to half of the cases in the 2000s. By contrast, the No Ponzi Game condition $r > g_D$ peaked in the 1960s (52%) and in the 1980s (54%), along with lower frequencies in the 1970s and the 1980s (35% and 40%). Both conditions simultaneously hold with a negligible proportion in the 1960s and the 1970s (at most 9%), peaked in the 1980s (53%), and declined in the 1990s (39%) and the 2000s (20%). Increases of the public debt over nominal GDP ratio where less frequent in the 1960s and 1970s (23% and 39%) and much more frequent in the three decades which followed (60%, 56%, 54%).

The most frequent regimes in the 1960s and the 1970s were regimes 5 and 4 (consolidation of public debt with respect to nominal output, with relatively low interest rate) and regime 1 (increase of the public debt/nominal output ratio with low interest rate, which represents 25% of the cases in the 1970s).

By contrast, the most frequent regimes in the 1980s and the 1990s was regime 2 ($g_D > r > g_Y$) for 41% of the cases, with an increase of public debt/nominal output ratio, and an interest rate larger than the growth rate of output. Then follows the transversality conditions regime 6 with consolidation (35% and 28% of the cases) and regime 3 with increases of public debt/nominal GDP (18% and 11%).

Finally, the 2000s presents a balanced breakdown of frequencies, with most frequent regimes being for increases of public debt over nominal GDP: regime 2 ($g_D > r > g_Y$, 29% of the cases), and regime 1 ($g_D > g_Y > r$, 21% of the cases). For consolidation of the public debt over nominal GDP, the most prevalent regimes are regime 5 ($g_Y > r > g_D$, 25% of the cases), and regime 6 (No Ponzi Game, transversality condition on GDP: 16% of the cases).

As a consequence, those "reference" textbooks based on both conditions distort the judgement of graduate students leading them to believe that **the high real interest rate regime of the 1980s decade** is a permanent long term regime. The 1980s were the decade where those conditions became a standard assumption in textbooks such as Blanchard and Fischer (1989).

## 5. Conclusion

An economic world where the No-Ponzi Game and the transversality conditions are always valid, as it may happen in contemporary reference macroeconomic textbooks on hundreds of pages, may not reflect what happened in the OECD countries over the period 1961-2010. They may mostly fit with recessions and the 1980's and 1990's decades. Hence, the doubts expressed by Blanchard and Weil (1992) related to the no Ponzi game condition and the real world upon the period 1960-1990 are still valid twenty years later upon the whole period 1960-2010. Finally, the claim that the No Ponzi game condition and the transversality condition insure solvency and debt/GDP consolidation is not validated by the data which are more in line with the Keynesian framework.

# References

Amable B. and Azizi K. (2011) "Varieties of capitalism and varieties of macroeconomic policy. Are some economies more procyclical than others?", Max Planck Institute Discussion paper 11/6.

Amable B., Chatelain J.B., and Ralf K. (2010) "Patents as collateral". *Journal of Economic Dynamics and Control.* vol. 34(6), pp. 1092 – 1104.

Arestis, Philip and Malcolm Sawyer (2004a) "On the Effectiveness of Monetary Policy and of Fiscal Policy", *Review of Social Economy*, December, LXII (4), pp. 441-463.

Arestis, Philip and Malcolm Sawyer (2004b) "On fiscal policy and budget deficits" *Intervention Journal of Economics*, 1(2), 65-78.

Barro R.J. (1974) "Are government bonds net wealth?" *Journal of Political Economy*, vol. 82(6), pp. 1095 – 1117.

Barro R.J. and Sala-I-Martin (2004) *Economic Growth* (2$^{nd}$ edition), MIT Press, Cambridge.

Blanchard O.J. and Fischer S. (1989). *Lectures in Macroeconomics.* MIT Press. Cambridge, USA.

Blanchard O.J. and Leigh S. (2013). Growth Forecast Errors and Fiscal Multipliers, IMF working paper, 13/1.

Blanchard O.J. and P. Weil: (1992): Dynamic Efficiency, the Riskless Rate, and Debt Ponzi Games under Uncertainty, NBER working paper (later published in *the Berkeley Electronic Journals in Macroeconomics*, *Advances in Macroeconomics*: 2001).

Caputo M.R. (2005): *Foundations of Dynamic Economic Analysis: Optimal Control Theory and Applications.* Cambridge University Press. Cambridge, United Kingdom.

Chatelain J.B. (2010): Can Statistics Do Without Artefacts? Cournot Centre, Prisme 19.

Chatelain J.B. and Ralf K. (2012a): The Failure of Financial Macroeconomics and What to Do about It. *The Manchester School.* 80 (S1), pp.21-53.

Chatelain J.B. and Ralf K. (2012b). Spurious regression and Near Multicollinearity, with an application to Aid Policies and Growth. Centre d'Economie de la Sorbonne working paper 12/078.

Chiang, A.C. (1992), *Elements of Dynamic Optimization,* New York: McGraw-Hill, Inc.

Domar E. (1944) "The 'Burden of the Debt' and the National Income", *American Economic Review*, December, 34 (4), pp. 798-827.

Hansen A. and G. Greer (1942) "The Federal Debt and the Future", *Haper's Magazine*, avril, pp. 489-500.

Haliassos M. and Tobin J. (1990). "The Macroeconomics of Government Finance". In *Handbook of Monetary Economics*, vol. 2, (B.M. Friedman and F.H. Hahn editors), Elsevier Science Publishers, Amsterdam. 889-959.

Halkin, H. (1974), "Necessary Conditions for Optimal Control Problems with Infinite Horizons," *Econometrica*, 42, 267–272.

Heijdra B.J. and Van Der Ploeg F. (2002). *Foundations of Modern Macroeconomics.* Oxford University Press, Oxford.

Herndon T, Ash M. and Pollin R. (2013). "Does High Public Debt Consistently Stifle Economic Growth? A Critique of Reinhart and Rogoff", PERI working paper, University of Massachusetts, Amherst.

Kamihigashi T. (2005). "Necessity of the Transversality Condition for Stochastic Models with Bounded or CRRA Utility," *Journal of Economic Dynamics and Control* 29(8), 1313-1329.

Keynes J.M. (1923). "A Tract on Monetary Reform". London, Macmillan.

Lerner A. (1943) "Functional finance and the federal debt", *Social Research*, 10:1/4, pp. 38-51.

Michel, P. (1982), "On the Transversality Condition in Infinite Horizon Optimal Problems," *Econometrica*, 50, 975–985.

Piketty T. (2011) "On the long run evolution of inheritance: France 1820-2050." *Quarterly Journal of Economics,* 126, 1071-1131.

Pucci M. and Tinel B. (2011) "Réductions d'impôts et dette publique en France », *Revue de l'OFCE*, 116, janvier, pp. 125-148.

Ramsey F.P. (1928) «A mathematical theory of saving », *The Economic Journal*, 38(152), pp. 543-559.